# Towards an automated repository for indexing, analysis and characterization of municipal e-government websites in Mexico

**Sergio R. Coria[1], Leonardo Marcos-Santiago[2], Christian A. Cruz-Meléndez[3], Juan M. Jiménez-Canseco**

[1] (Corresponding author) Research Professor, Institute of Informatics, and Graduate Program in e-Government, University of the Sierra Sur (UNSIS).
Postal address: Calle Guillermo Rojas Mijangos S/N, Esq. Av. Universidad, Col. Ciudad Universitaria, C.P. 70800, Miahuatlán de Porfirio Díaz, Oax., Mexico
E-mail address: srco2001@yahoo.com and coria@unsis.edu.mx

[2] Undergraduate student, School of Informatics, UNSIS.

[3] (Cátedra CONACyT) Research Professor, Graduate Program in e-Government, UNSIS.

Declarations of interest: none

This research did not receive any specific grant from funding agencies in the public, commercial, or not-for-profit sectors.

## Abstract

This article addresses a problem in the electronic government discipline with special interest in Mexico: the need for a concentrated and updated information source about municipal e-government websites. One reason for this is the inexistence of a complete and updated database containing the electronic addresses (web domain names) of the municipal governments having a website. Due to diverse causes, not all the Mexican municipalities have one, and a number of those having it do not present information corresponding to the current governments but, instead, to other previous ones. The scarce official lists of municipal websites are not updated with the sufficient frequency, and manually determining which municipalities have an operating and valid website in a given moment is a time-consuming process. Besides, website contents do not always comply with legal requirements and are considerably heterogeneous. In turn, the evolution development level of municipal websites is valuable information that can be harnessed for diverse theoretical and practical purposes in the public administration field. Obtaining all these pieces of information requires website content analysis. Therefore, this article investigates the need for and the feasibility to automate implementation and updating of a digital repository to perform diverse analyses of these websites. Its technological feasibility is addressed by means of a literature review about web scraping and by proposing a preliminary manual methodology. This takes into account known, proven, techniques and software tools for web crawling and scraping. No new techniques for crawling or scraping are proposed because the existing ones satisfy the current needs. Finally, software requirements are specified in order to automate the creation, updating, indexing, and analyses of the repository.

**Key words:** electronic government, local government, municipal *websites*, descriptive statistics, web data mining, software engineering, Mexican municipalities, State of Oaxaca.

## 1. Introduction

In Mexico, alike other developing countries, e-government started in the late 1990s and early 2000s. One of its first implementations was the set of websites of the Federal Executive Branch that were propitiated by the Mexico Online, and e-Mexico programs (UNESCO, 2002:56-60). In the following years, also the State governments in this country started creating their websites; e.g. the State of Mexico (UNESCO, 2002:59) and, finally, a number of municipalities did it, too (Gómez-Reynoso and Sandoval-Almazán, 2013).

e-Government websites are usually analyzed and characterized on the basis of their evolution development level; i.e. how mature they are, considering their possibilities to satisfy the needs of citizens and also those of a government in order to produce public value. This characterization is mostly based on the general lines stated by the *Web Presence Measurement Model* (UN, 2003:12). This index quantifies how much a government website is used as a tool to inform, interact, transact, and network. The process to characterize a website involves that an observer identifies content items and services (or functionalities) in the website to determine its corresponding evolution development level. The analysis and characterization process is a manual/visual task; however, it can be automated (or semi-automated) by using software tools. Therefore, this research aims at exploring the need and feasibility to automate the search, analysis, and characterization of the municipal websites in Mexico. Based on all of the above, this article is organized as follows: Section 2 presents the problem definition. Section 3 presents the background on e-government websites. Section 4 briefly reviews works addressing the analysis and characterization of municipal government websites, and web scraping techniques and tools. Section 5 describes our methodology. Section 6 presents preliminary results. Section 7 presents software requirements specifications for implementing the repository, and, finally, section 8 presents conclusions and suggests future research work.

## 2.   Problem definition

Mexico is a federal republic organized into 32 States that, in turn, are organized into municipalities. The total number of municipalities in the country is 2,457[1] (including the 16 *delegaciones* that constitute the political division in Mexico City). In early 2018, in Mexico, e-government is a phenomenon whose rising and development at the federal level are higher than those at the State and municipal levels. Regarding the governments of the 32 States, significant differences exist among the evolution development levels of their websites from a public administration perspective (Sandoval-Almazán, 2015). However, similarities exist among them on the basis of the contents organization into sections. For instance, most of the State websites have sections entitled as: *transparency*, *tourism*, *procedures and services*, etc. In turn, the municipal government websites present a high heterogeneity. A first issue is the technical difficulty to know which municipalities have an official website corresponding to its current government. Nowadays, there is no official directory containing all the updated Internet addresses (domain names, or uniform resource locators) of these websites. In addition, eventually, the official domain name of any municipal website can be in an unusual status due to technological or administrative reasons that limit, obstruct or completely impede visualizing the website. For instance: a domain name can be in suspension status due to non-payment of name usage, its setting parameters on domain name servers (DNS) can be incorrectly configured, etc. Moreover, a municipal website can contain obsolete information corresponding to any previous government and not to the current one.

---

[1] http://cuentame.inegi.org.mx/territorio/division/default.aspx?tema=T

Manually producing and updating a directory that can contain the official domain names of all Mexican municipalities at a given time is a difficult and time-consuming process. In addition to the web search tools (i.e. Google, Bing, etc.), there are a small number of information sources to find the municipal websites. However, they are not complete and updated on a sufficiently regular basis. One of these sources is the Mexican Institute for Federalism and Municipal Development (INAFED)[2]. This has developed the Mexican System on Municipal Information (SNIM)[3], offering a downloadable dataset that includes the domain names. Nevertheless, the working status, validity and availability of the websites in this database are uncertain. Although INAFED is in charge of a series of databases with municipalities' information, its updating frequency does not suffice.

An alternate solution to find the official websites could be by using the information and services of Akky[4], a Mexican company working as the main *registrar* of domain names with the *.mx* suffix (including *.gob.mx*). Nevertheless, although this company could produce a list or a directory with all the *.gob.mx* domain names, at the present time it does not offer an updated directory specifically containing the municipal domain names. In addition, the municipal *websites* are usually modified when a local government is replaced by the new authorities after elections. These modifications can involve changes in contents and even in the domain name. Therefore, our research aims not only at solving this problem, but, besides, at providing other additional benefits by concentrating a digital repository with automatic functionalities for contents analysis and classification.

---

[2] https://www.gob.mx/inafed (accessed: October 17, 2017).
[3] http://www.snim.rami.gob.mx (accessed: October 17, 2017).
[4] https://www.akky.mx/ (accessed: October 17, 2017)

In international scientific literature, three of the most frequent analyses on e-government websites (either national or subnational) are: 1) verifying that their contents and services comply with legal requirements (e.g. on transparency), 2) characterizing their evolution development level according to particular theories, and 3) evaluating their usability. In the first type of analyses, the purpose is to verify the presence of certain contents and website functionalities; e.g. information on: financial incomes and expenses, salaries of government officials, tenders and award of contracts, website functionalities for consulting and paying taxes, etc. In the second type of analyses, the contents and functionalities that are identified in the website are used to classify it into a class. This is one among a set of classes that are previously defined on a theoretical basis addressing the evolution development of government websites. A number of these theories have been proposed; for instance, (Layne and Lee, 2001; Srivastava and Teo, 2004; Gil-García and Martínez-Moyano, 2007; Sandoval-Almazán y Gil-García, 2009; Fath-Allah et al., 2014, among others.). Usually, the classes or levels of evolution development of e-government websites are enumerated as: presence, informative, interactive, transactional, integrative, participative, etc. Finally, the third type of analyses (on usability) consists in measuring (either quantitatively or qualitatively) how usable a website is; i.e. how easy or how difficult its usage is for non-expert users (e.g. Youngblood and Mackiewicz, 2012; Meijerink, 2016; Martínez-López, 2016).

Previous research work on e-government in Mexico has been mainly focused on the federal and State levels (see *Section 4. Literature review*). Regarding the municipal level, research works are highly scarce. A relevant theoretical and practical problem in this matter is the scarcity of information about which Mexican municipalities have an official e-government website in a given time. Also, there are few studies analyzing the general contents of these websites; i.e. how the content is organized or what are the most frequent sections in them. Finally, determining their evolution development level is a task that can be automated. Considering all of the above, there exists a need for methods and automatic tools to find, analyze, and characterize the official, valid, e-government websites of the Mexican municipalities. This is the research problem that is addressed by this work from the perspectives of digital government and computer science.

## 2.1. Subject matter

For the purposes of our research, the subject matter is the set of official e-government websites of the approximately 2,457 Mexican municipalities, considering that a relatively large (and hard to quantify) number of these do not have an official website. In this research, an official municipal website is that whose domain name (*i.e.* its address in the World Wide Web) contains the suffix *.gob.mx*. It does not matter if the beginning of this name includes the prefix *www* or not. Instances of official domain names are: *www.municipiodeoaxaca.gob.mx*, *municipiomiahuatlan.gob.mx*, *www.sanjoselachiguiri.gob.mx*, etc. In contrast, domain names containing any other suffixes different from *.gob.mx*, such as *.com*, *.com.mx*, *.org*, *.org.mx*, *.net*, *.net.mx*, etc. are not considered official by the Mexican regulations and, thus, are discarded from this research. Instances of these names are: *oaxaca.com*, *huatulco.com.mx*, *salinacruz.com*. Official Facebook profiles of the municipal governments are also discarded.

## 2.2.  Research assumption, approach and scope

No hypothesis is proposed; however, our main assumption is that the diverse analyses and characterizations that use to be performed on the contents of municipal websites (e.g. compliance with legal requirements, characterization of their evolution development level, etc.) can be automated. This assumption involves leveraging data science methods and software tools to automate tasks such as web crawling, web scraping, and descriptive statistical analyses. The approach is mainly quantitative and the scopes are descriptive and exploratory because, to our knowledge: 1) there are no comprehensive enough and updated sources of information about the Mexican municipal websites, and 2) there is a small number of previous research works on this topic. With this in mind, implementing the digital repository is not within the scope of this research, but instead, exploring its need and its technological feasibility for a subsequent implementation.

## 2.3. Research objectives

The overall research objective is: *to analyze the need and the technological feasibility to automate the creation, indexing, and analyses of a digital repository containing replicas of official e-government websites of the Mexican municipalities.* This is relevant because: 1) in 2018, there are no government, private, academic or social sources on the WWW that concentrate a complete and updated list of these websites; 2) the relative large number of municipalities in Mexico and the renewal of their authorities after local elections involve a persistent lack, deficiency or heterogeneity of information in the websites, 3) identifying the existence and the evolution development level of these websites is highly useful for promoting the e-government benefits in this country, and 4) diverse information technologies are available and can be harnessed to automatically or semi-automatically produce directories, databases, and analyses from websites. In order to achieve the general objective, specific objectives need to be achieved. These are as follows:

i. Determine how to find the official e-government websites of all the Mexican municipalities

ii. Distinguish between the websites corresponding to valid (current) municipal governments and those to previous ones

iii. Manually identify the most frequent contents in the websites in order to specify subsequent automated analyses and characterizations

iv. Propose a technique to automate the classification of the evolution development level of the websites

v. Specify general software functionality requirements to automate the extraction of specialized information and the production of statistical analyses (cross-sectional or synchronic, and longitudinal or diachronic) and characterizations on the websites.

## 2.4. Research justification

Implementing a digital repository of replicas of the Mexican municipal e-government websites for analysis and classification is a relevant and original problem and a technologically feasible task. The problem is relevant in the electronic government and the computer science disciplines because: 1) it is complex, significant and underexplored, 2) websites are the main means in e-government for communication with citizens and they are frequently studied by theorists and practitioners in diverse knowledge fields, 3) the addressed problem involves links with the potential benefits of e-government to society, such as increasing transparency, accountability, and citizen participation, and 4) computer science can offer a series of methods and tools to facilitate and even to automate the study of these websites. The expected results in this research will be useful to the academic, government, corporate, and social sectors due to the reasons that are enumerated next. From an academic perspective, there exist scarce works addressing municipal e-government in Mexico. Our results will provide researchers in the public administration, e-government, informatics, and computer science areas with useful information resources, methods and tools to perform studies on: compliance of legal requirements in the websites, transparency, web mining, software engineering, etc. Diverse analyses on the websites contents can be performed to identify similarities and differences among them. Moreover, a potential application is to automate the production of longitudinal studies to analyze all these websites along broad periods, beyond the duration of specific municipal governments. In general terms, new phenomena and research problems can be identified and studied by means of an information resource such as this repository.

From the government perspective, a frequently updated and complete directory of valid municipal websites can be highly useful for the purposes of public administration in general. The findings from the websites analyses can provide insights to create or modify laws or technical standards on contents, structure, or services in these websites. Also, governments can obtain insight regarding the evolution development of these websites.

Private companies in the information technology sector can obtain useful information to identify business opportunities for developing and offering products or services to local governments in order to improve contents, functionalities and usability of municipal websites. In addition, companies in other diverse sectors can find potential government clients to offer products and services such as infrastructure construction or government consulting. Mexican citizens, in general, can use these websites to leverage information and services from municipal governments, increasing the civic engagement. Finally, this research is feasible because the pieces of the subject matter (*i.e.* the municipal websites) are freely available and there are a number of web crawling and scraping techniques and free software tools that can be harnessed for research purposes.

## 3.  Background

Sandoval and Gil-García (2009) comment that there exist difficulties, such as the fast technological innovation, political changes and the brief continuity in public policies that impact on e-government websites evolution. In the particular case of Mexico, there is no regulation on electronic government that obligates to uniformity or continuity in websites after the replacement of public administrations after elections. In general terms, there are no integral public policies on e-government. Therefore, there are no evaluation models for e-government websites that are uniform or that consider all the involved components. Thus, these analyses need to consider the context in which a website is developed and also its objectives. These can be based on the interests of the public administration or on citizens' demands.

A series of models have been produced to analyze and characterize the evolution development level of e-government websites. These usually define development stages that aim at comprehend the components and scope of the websites. (Gil-García and Martínez-Moyano, 2007) call them *evolution models*, involving that electronic government, in general, and e-government websites, in particular, are constantly evolving and adding technological and organizational sophistication. In turn, (Fath-Allah et al., 2014) present a review of approximately 25 different *electronic government maturity models*, considering a model as a set of stages (from basic to advanced) that determine the maturity of an e-government website. Two key benefits of evolution or maturity models are: 1) they offer a manner to classify the e-government portals, and 2) they can be useful to agencies for enhancing their websites quality. Since 2003, the United Nations (UN, 2003) assess the evolution development of the national e-government websites of its member countries. This evaluation is performed on the basis of a four-stage model: 1, websites with emerging information services; 2, websites with enhanced information services; 3, websites with transactional services; and 4, websites with connected services.

## 4. Literature review on municipal website analysis and web scraping

Our review of scientific literature is focused on two main topics: 1) municipal website analysis (see Table 1), and 2) web scraping (see Table 2). The purposes of this review are to obtain an overview of the most frequent types of analyses that are performed on the municipal websites contents worldwide, and of the most used techniques and free software tools that are already available and can be used to automate the identification of the evolution development level of these websites. The review supports our research to build on previous works to introduce the potentially innovative notion of *repository of municipal e-government websites*. In addition, this review allows identifying techniques and tools for automating the classification of these websites into categories of evolution development levels. The purposes of this review are mostly practical and less theoretical, with the objective to guide the repository implementation. However, once the repository is implemented, it can be used to discover novel phenomena or problems in the e-government theory and practice and in its social impacts.

From an international perspective (not including Mexico), a number of research works analyzing municipal e-government websites exist. Table 1 enumerates a collection of them, which analyze aspects such as: evolution development, usability, credibility, citizen participation, user engagement, accessibility for disabled users, etc. A subset of the most interesting works is that by Holzer et al. (2003, 2005, 2007, 2009, 2012, 2014 and 2016). This series analyzes municipal websites of a number of cities in diverse countries as a longitudinal study. In turn, (Huang, 2007) is remarkable because analyzes the largest number of websites in our literature review: 1,744 counties in U.S.A. Moreno-Sardá et al. (2013) study 947 websites in Catalonia, Spain. Feeney and Brown (2017) analyze 500 municipalities in U.S.A. Finally, Piñeiro-Naval et al. (2017) study 500 websites in Spain.

To our knowledge, previous research works addressing the existence, content structure, or evolution development of the Mexican municipal websites are scarce. Most of research on e-government in Mexico is focused on federal or State websites. A series of the most known works are the annual rankings of websites of the 32 State governments since, approximately, 2006 to 2015 presented in: Luna-Reyes, L.E. et al. (2011), Luna-Reyes, D.E. et al. (2011), Sandoval-Almazán (2015), Sandoval-Almazán and Gil-García (2008), Sandoval-Almazán et al. (2010), and Sandoval-Almazán et al. (2012). These analyses are mostly based on methodologies presented in (Gil-García et al., 2007), (Gil-García and Martínez-Moyano, 2007), (Sandoval-Almazán y Gil-García, 2008), and (Sandoval-Almazán y Gil-García, 2009).

One of the scarce works addressing municipal websites in Mexico is (Sandoval-Almazán and Mendoza-Colín, 2011). This analyzes a sample of 518 (21.1%) out of the 2,454 municipalities in this country in 2010. Its results reveal a poor usage of ICT in most of them. Also, deficiencies are identified in aspects such as: open government, web design, usage of Web 2.0 tools, and applications for interaction between municipal governments and citizens. One of their main conclusions is that: "the Mexican municipalities are in the initial stages of the e-government development and most of them do not have the infrastructure and resources necessary to implement it". In turn, Martínez-López (2016) evaluates the usability of the municipal website of Miahuatlán de Porfirio Díaz, in the State of Oaxaca.

**Table 1.** Works on analysis of municipal e-government websites (not including Mexico).

| No. | Reference | Continent, country or province | Analyzed aspects | Number of analyzed websites |
|---|---|---|---|---|
| 1 | Cardoso de Miranda, and Muñoz-Cañavate (2015) | Portugal (all municipalities in 2010) | Services on: information, communication, and transactions | 308 |
| 2 | Dolson and Young (2012) | Canada (municipalities between 20,000 and 125,000 inhabitants, except those in Quebec) | e-Content, e-Participation, and social media capacity | 109 |
| 3 | Feeney and Brown (2017) | U.S.A. (municipalities between 25,000 and 250,000 inhabitants) | Information tools, e-services, utility, transparency, and civic engagement | 500 |
| 4 | Holzer et al. (2003, 2005, 2007, 2009, 2012, 2014, 2016) | Worldwide (the most populated city in countries with the highest percentages of Internet users) | Privacy and security policy, usability, content, services, citizen participation | 84, 81, 86, 87, 92, 100, 97, respectively |
| 5 | Huang (2007) | U.S.A. (all counties official websites) | Information, communication, transaction and democracy | 1,744 |
| 6 | Moreno-Sardá et al. (2013) | Spain (Catalonia) | Availability of specific information items | 947 |
| 7 | Piñeiro-Naval et al. (2017) | Spain (a sample of municipalities) | Visual appearance, information architecture, and usability | 500 |
| 8 | Wohlers (2008) | U.S.A. (municipalities in 5 States) | Interactive democracy, service delivery, and specific information items | 200 |

Source: own elaboration using information from the respective authors.

**Table 2.** Brief review of literature on techniques and software tools for web scraping.

| No. | Year | Authors | Highlights |
|---|---|---|---|
| 1 | 2000 | Craven et al. | A methodology to construct knowledge bases from the World Wide Web |
| 2 | 2004 | De Castro Reis et al. | Automatic scraping and a tree edit distance algorithm are used for news extraction from the web |
| 3 | 2007 | Liu | Textbook on web data mining that includes scraping techniques |
| 4 | 2007 | Markov and Larose | Textbook on web data mining that includes scraping techniques |
| 5 | 2007 | Christos et al. | A web content manipulation technique based on page fragmentation to produce personalized views of websites |
| 6 | 2010 | Yang et al. | An automated climatic data scraping, filtering and display system |
| 7 | 2012 | Chernysh | (U.S. Patent) Creation of data extraction rules to facilitate scraping of unstructured data from web pages |
| 8 | 2012-2018 | Richardson | *BeautifulSoup*, a package for scraping in Python language |
| 9 | 2013 | Glez-Peña et al. | A description of two bioinformatics meta-servers that use scraping for gene set enrichment analysis |
| 10 | 2014 | Nair | Textbook about the *BeautifulSoup* Python package |
| 11 | 2015 | Sirisuriya | Review on scraping techniques and software tools |
| 12 | 2015 | Munzer et al. | Textbook on automated data collection using the *R* language for web scraping and text mining |
| 13 | 2015 | Polidoro et al. | Techniques to collect data on consumer electronics and airfares |
| 14 | 2015 | Lawson | Textbook on web scraping using the Python language and its packages |
| 15 | 2015 | Bonifacio et al. | A free tool for scraping Canadian climate data |
| 16 | 2015 | Mitchell | Textbook on web scraping using the Python language |
| 17 | 2016 | Boeing and Waddell | Scraping is applied to analyze listings on rental housing markets |
| 18 | 2016 | Turland | A textbook for web scraping using the *PHP* language |
| 19 | 2016 | Ozacar | A tool for producing structured interoperable data from product features on the web |
| 20 | 2016 | Massimino | Web crawling and information scraping techniques to automate the assembly of research data in the business logistics field |
| 21 | 2017 | Khalil and Fakir | They present *Rcrawler*, a library for the *R* software that crawls and scrapes web sites |
| 22 | 2017 | Bougrine et al. | Metadata associated to multi-media resources are scraped by using the *BeautifulSoup* Python package |
| 23 | 2017 | Wimmer and Yoon | Web scraping, natural language processing, and other methods are applied to analyze customer reviews of products in online commerce |
| 24 | 2018 | Krotov and Tennyson | Financial data are scraped from the Web using packages in *R* language |
| 25 | 2018 | Slamet et al. | Methodology for scraping and naïve Bayes classification in a job search engine |

**Source:** own elaboration using information from the respective authors.

## 5. Proposed methodology

This research has a quantitative, descriptive and exploratory approach. Unlike the identified related works, both in Mexico and other countries, one of the ultimate goals of our methodology is automating analyses and characterizations of the municipal websites by means of statistical methods and software tools that are available in the web data mining discipline (Liu, 2007). In a subsequent stage of this research, the methodology will allow the automatic production and periodical updating of a digital repository of replicas of the Mexican municipal websites. The repository will allow automating a series of tasks, such as: 1) a highly frequent updating of a list (directory) of municipal websites, 2) the extraction and organized storing of useful textual (or non-textual) information to produce specialized directories or databases, e.g. contact data, photographs, maps, etc., and 3) characterization of the evolution development level of the websites. For these reasons, the presented methodology is based on non-automated or semi-automated activities that are considered for future automation. The methodology is mainly focused on producing a list of the municipal e-government websites and a list of the content sections in each of them. Its general steps are described below.

## 5.1. Find the official domain names of the municipal e-government websites in Mexico and identify their operating status

The official domain names of municipal e-government websites (*i.e.* only those with the *.gob.m*x suffix, discarding others with any other different suffixes) can be found by using one or more of the following resources:

i. The database produced by the Mexican Institute for the Federalism and Municipal Development (INAFED), available at the Mexican System on Municipal Information[5].

ii. The services for consultation and purchasing of *.mx* domain names provided by Akky (https://www.akky.mx), the private company that is in charge for administering domain names with this country suffix.

---

[5] http://www.snim.rami.gob.mx (accessed: October 17, 2017)

iii. Performing a semi-automatic searching of the websites by means of crawling software tools, also known as *crawlers* or *spiders*; for instance, the J-Spider freeware[6]. A crawler can automatically verify the existence (and other characteristics) of a website (or set of websites) by feeding the tool with a domain name or a set of domain names. Also, online services for crawling, such as *Octoparse*[7] can be used for this purpose.

iv. Manual searching of the domain names by means of Google, Bing, Twitter or Facebook (but not considering Facebook accounts of municipalities as official websites).

v. Once the official websites are found, an additional, simple, task can be performed for each website: identifying its hosting provider and its location country. This information is useful to discover potential concentration of hosting services in specific providers. The task can be performed by using tools such as *Who is Hosting This*[8], among others.

The main output of these activities is a table of municipalities with information on:

i. Municipality name.

i. The official domain name that was found and its corresponding operating status. The status can be: *working* (i.e. the domain name is associated to any web page), or *not working*. (*i.e.* the domain name is not associated to any web page).

ii. Date of access; i.e. the date in which the website was consulted. This is a useful data because municipal websites can change its working status or its content on a non-regular basis.

iii. The name of the web hosting provider and its location country.

iv. The above mentioned data are complemented with the municipality ID. This is obtained from the Municipal Geo-Statistical Framework[9] that is produced by the Mexican Institute for Statistics and Geography (INEGI). This is an integer number identifying the municipality. It is broadly used by INEGI to associate diverse geographic, demographic and economic data and maps facilitating consultation, data joining and analyses.

---

[6] http://j-spider.sourceforge.net/ (accessed: May 31, 2017).
[7] http://www.octoparse.com/ (accessed: May 31, 2017).
[8] https://www.whoishostingthis.com/ (accessed: May 31, 2017).

In order to determine the completeness of the domain name list, these conditions must be accomplished:

i. All Mexican municipalities should have, at least, one domain name in the list. This can be verified by comparing to a list of all municipality names obtained from INEGI

ii. All domain names in the list have to be in a valid status, i.e. the domain names must be associated to any web content different from messages of domain name suspension

However, exceptions to these conditions are: municipalities for which a domain name cannot be found after manual, semi-automatic, or automatic search in the Web.

## 5.2. Identify the government periods of the websites

In every official website whose status is *working*, the initial and final years of the period corresponding to the municipal government responsible for the website content are identified. This information is necessary to determine if the website corresponds to the current municipal government or to any previous one. This task can be automated or semi-automated by using *scraping* software tools, such as *Google Web Scraper*[10], *Octoparse*, among others. The government period is incorporated into the table containing the municipality name, the official domain name, etc.

## 5.3. Produce statistical analyses of information related to the domain names

Once the table is completed, two types of statistical analyses are produced on its data: 1) Pareto tables, and 2) choropleth maps. Pareto analyses are produced to find the most and the least frequent cases on: working status, government periods, web hosting providers, and location countries of these providers.

---

[9] http://www.inegi.org.mx/geo/contenidos/geoestadistica/m_geoestadistico.aspx (accessed: May 31, 2017).
[10] https://chrome.google.com/webstore/detail/web-scraper/jnhgnonknehpejjnehehllkliplmbmhn?hl=en (accessed: January 17, 2018)

Choropleth maps for States (by using gray tone or color scales) are produced to visualize: municipalities with or without an official domain name, working status of the existing official domain names, and government periods of the websites.

**5.4. Identify and analyze the general contents in the websites**

Most of the Mexican municipal websites do not present their information organized into standardized sections or subsections. Therefore, a general exploratory approach for a preliminary analysis of their contents is by identifying and counting the titles of the major sections in them. These sections are presented at the highest hierarchy menu that is located in the main homepage (the *start* page) of each website. Thus, a list of section titles is produced from that menu in every website. At the present time, scraping software tools (e.g. Google Web Scraper, Octoparse, etc.) can be useful to automate or semi-automate the search for the content titles in the websites.

Once the list with section titles is generated, Pareto analyses are produced for: number of sections in the websites, number of websites per number of sections (i.e. how many websites have certain number of sections), and number of websites containing certain section titles (e.g. how many websites contain a section entitled *Transparency*, or *Tourism*, or *Procedures and services*, etc.). No deeper analyses are performed on the websites at the present stage of this research due to the potentially large number of websites in this country: 2,457 municipalities with an unidentified number of working, valid, official websites. However, these exploratory analyses identify, in general terms, the most frequent contents and obtain an overview of the municipal e-government phenomenon in Mexico. Subsequent stages of this research will identify and quantitatively analyze more specific characteristics and contents in the websites.

**5.5 Classify the evolution development level of the websites by using web scraping and cue words or phrases**

The classes of evolution development levels of the e-government websites are mutually exclusive because the higher levels have functionalities that are not present in the lower ones. For instance, a website in the *participation* level has functionalities that are not present in *transaction* or *interaction* or *information* level websites. In turn, a *transaction* level website has functionalities that are not present in its two preceding levels, etc. Typical functionalities of each level are usually associated to certain cue words or phrases.

Therefore, the automatic classification of municipal websites into a set of classes of evolution development level can potentially be implemented as an automatic text classification task on the basis of cue words or phrases that are frequently present in the HTML source code of websites corresponding to each level. For instance: *participation* level can be recognized by finding words such as: *participative budget*, etc. *Transaction* level can be identified by words such as: *pay*, *tax*, *receipt*, etc. *Interaction* can be recognized by finding words such as: *online services*, *request*, *consult*, etc. Finally, *information* level can be recognized by the absence of cue words or phrases corresponding to the higher development levels. For this purposes, regular expressions can be used into scraping freeware tools or in *ad hoc* scripts programmed in languages such as Python, R, or PERL. Table 3 presents potential cue words/phrases corresponding to each development level. These words/phrases are obtained from the statistical analysis of section titles in municipal websites of the State of Oaxaca (see Table 5 in Section 6.1 *Statistical results*) and also by empirical observation on websites of other States (see Figures 2 through 5).

**Table 3.** Potential cue words or phrases in municipal websites in each evolution development level.

| No. | Evolution development level | Potential cue words or phrases (originally in Spanish) | Excluding cue words or phrases |
|---|---|---|---|
| 1 | Information | Transparency, telephone, fax, e-mail, @, social networks, follow us, Facebook, Twitter, Instagram, YouTube | Those corresponding to levels 2, 3 or 4 |
| 2 | Interaction | Online services, complaints, suggestions, requests, follow-up, report, chat, consultation, consult, online consult, download | Those corresponding to levels 3 or 4 |
| 3 | Transaction | Payments, online payments, pay, payment, tax, property tax, receipt, penalty fee (fine, ticket), bill, invoice, property tax consultation, property tax payment. | Those corresponding to level 4 |
| 4 | Participation | Participative budget, participative, opine. | *** |

**Source:** own elaboration.

As website classes are mutually exclusive (*i.e.* a website belongs to one class only), the automatic classification process of a website can be facilitated if it is performed as a discarding task, as described in a simple algorithm below:

   i. Start

   ii. ¿Does the website contain cue words/phrases of the *participation* level?

      Yes: classify as *participation*, go to step *v*.

      No: continue.

   iii. ¿Does the website contain cue words/phrases of the *transaction* level?

      Yes: classify as *transaction*, go to step *v*.

      No: continue.

   iv. ¿Does the website contain cue words/phrases of the *interaction* level?

      Yes: classify as *interaction*, go to step *v*.

      No: classify as *information*, go to step *v*.

   v. End

Defining a comprehensive set of the suitable cue words/phrases needs a statistical analysis on a sample of the Mexican municipal websites. This process can be facilitated if a digital repository of website replicas is implemented, although this is not essential for this specific purpose. Once an initial set of cue words/phrases associated to each evolution level is available, it can be gradually expanded as the repository stores a larger number of replicas. A sample size of websites has to be determined to perform this exploratory analysis. Once the most frequent cue words for each level are identified in a statistically representative sample of the websites, the classification process for other additional websites can be automated or semi-automated.

Once an initial website replica repository is implemented containing a relatively low number of websites, their evolution level can be manually determined and tagged (labeled, annotated) and their cue words/phrases can be identified. The repository can be used as an annotated corpus for research or practical purposes, including the automation (or semi-automation) of the classification of other additional websites. The initial collection of cue words associated to each development level can be used to classify other additional websites. In addition, the original set of cue words can be expanded by identifying new cue words/phrases. In a subsequent stage, the usefulness of artificial intelligence methods (e.g. from the natural language processing, and the machine learning areas) can be explored as a potential solution for producing an automatic classification model using sets of cue words/phrases as predictors and evolution level classes as target.

## 6. Preliminary results

The results are still preliminary because: 1) implementing and testing the repository is a time-consuming process that will be addressed in the next stage of the present research, and 2) approximately 1,800 municipalities in the other 31 States of Mexico should also be analyzed by using our methodology. These results are organized into two parts: first, a collection of tables, statistical charts and a map, and, then, an exploration of the technological feasibility for automating the classification of evolution development levels by using known techniques and tools for web scraping and cue words or phrases. In addition to the results, the subsequent section offers software requirements specifications to implement the automated repository of municipal website replicas with functionalities for information extraction, analyses, and characterization.

## 6.1. Statistical results

In order to test our methodology, the 570 municipalities in the State of Oaxaca are selected as a universe for an exploratory analysis due to these reasons: 1) the total number of Mexican municipalities (2,457) is too large to be processed at the present stage of this research, 2) Oaxaca is the Mexican State with the largest number of municipalities (approximately 23% of the national total), and 3) despite the significance of the number of municipalities in this State, in early 2018, there are no government, academic, commercial or social information sources providing an updated list of its municipal websites, so there exist a necessity and opportunity to produce it. Additionally, it is necessary to select a lower number of municipal websites from other States because most of municipal websites in the State of Oaxaca are in the lowest evolution development levels. Table 4 presents information of a subset from the 84 official domain names of municipal governments in the State of Oaxaca. These data include: INEGI ID (a numeric identifier defined by the Mexican Institute for Statistics and Geography), date of access to the website, municipal government period, and name and location country of the web hosting provider.

**Table 4.** Information of a subset of the 84 official domain names of municipal governments in the State of Oaxaca in late May, 2017 (we have uploaded the complete table at the Web[11]).

| No. | INEGI ID | Municipality name | Official web domain name | Date of access | Working | Government period | Web hosting provider | Provider's country |
|---|---|---|---|---|---|---|---|---|
| 1 | 002 | Acatlán de Pérez Figueroa | acatlandeperezfigueroa.gob.mx | 05/24/17 | Yes | Not specified | Servicios SSD | Spain |
| 2 | 005 | Asunción Ixtaltepec | asuncionixtaltepec.gob.mx | 05/24/17 | Yes | Not specified | LiquidWeb | USA |
| 3 | 009 | Ayotzintepec | ayotzintepec.gob.mx | 05/24/17 | Yes | 2014-2016 | Hosting Mexico | Mexico |
| 4 | 020 | Constancia del Rosario | www.constanciadelrosario.gob.mx | 05/24/17 | Yes | 2014-2016 | Microsoft Azure | USA |
| 5 | 030 | El Espinal | www.elespinal.gob.mx | 05/24/17 | Yes | 2017-2018 | Internet Names For Business.com | USA |
| 6 | 039 | Heroica Ciudad de Huajuapan de León | www.huajuapandeleon.gob.mx | 05/24/17 | Yes | 2017-2018 | GoDaddy | USA |
| 7 | 043 | Heroica Ciudad de Juchitán de Zaragoza | juchitandezaragoza.gob.mx | 05/24/17 | Yes | Not specified | Hosting Mexico | USA |
| 8 | 397 | Heroica Ciudad de Tlaxiaco | tlaxiaco.gob.mx | 05/28/17 | Yes | 2017-2018 | Colo4, LLC | USA |
| 9 | 041 | Huautla de Jiménez | www.huautladejimenez.gob.mx | 05/24/17 | Yes | 2017-2018 | BanaHosting | USA |
| 10 | 065 | Ixpantepec Nieves | www.ixpantepecnieves.gob.mx | 05/24/17 | Yes | 2014-2016 | Metro Net, S.A.P.I. de C.V. | Mexico |
| 11 | 044 | Loma Bonita | lomabonita.gob.mx | 05/24/17 | Yes | 2017-2018 | HostMonster | USA |
| 12 | 052 | Magdalena Tequisistlán | tequisistlan.gob.mx | 05/24/17 | Yes | 2014-2016 | Centrilogic, Inc. | USA |
| 13 | 562 | Magdalena Yodocono de Porfirio Díaz | municipioyodocono.gob.mx | 05/24/17 | Yes | 2014-2016 | HostGator | USA |
| 14 | 057 | Matías Romero Avendaño | municipiomatiasromero.gob.mx | 05/24/17 | Yes | 2017-2018 | HostGator | USA |
| 15 | 037 | Mesones Hidalgo | www.municipiomesoneshidalgo.gob.mx | 05/24/17 | Yes | Not specified | GoDaddy | USA |
| 16 | 059 | Miahuatlán de Porfirio Díaz | municipiomiahuatlan.gob.mx | 05/24/17 | No | Not specified | *** | *** |

Source: own elaboration using information from the municipal websites and www.whoishostingthis.com (accessed: May 24-28 and 31, 2017)

---

[11] https://drive.google.com/open?id=1m1ERFQDI6zzOovNX3HZcJhm0cd9JWupb

Figure 1 presents a map of the State of Oaxaca showing the condition of its municipal websites. The map allows visualizing that there exist a large number of municipalities that do not have an official domain name. Also, patterns of geographical groupings can be seen as sets of neighbor municipalities that have official domain names; for instance, clusters in Central Valleys (*Valles Centrales*), Coast (Costa), and Isthmus (*Istmo*).

**Figure 1.** Status of the municipal e-government websites in the State of Oaxaca in late May, 2017.

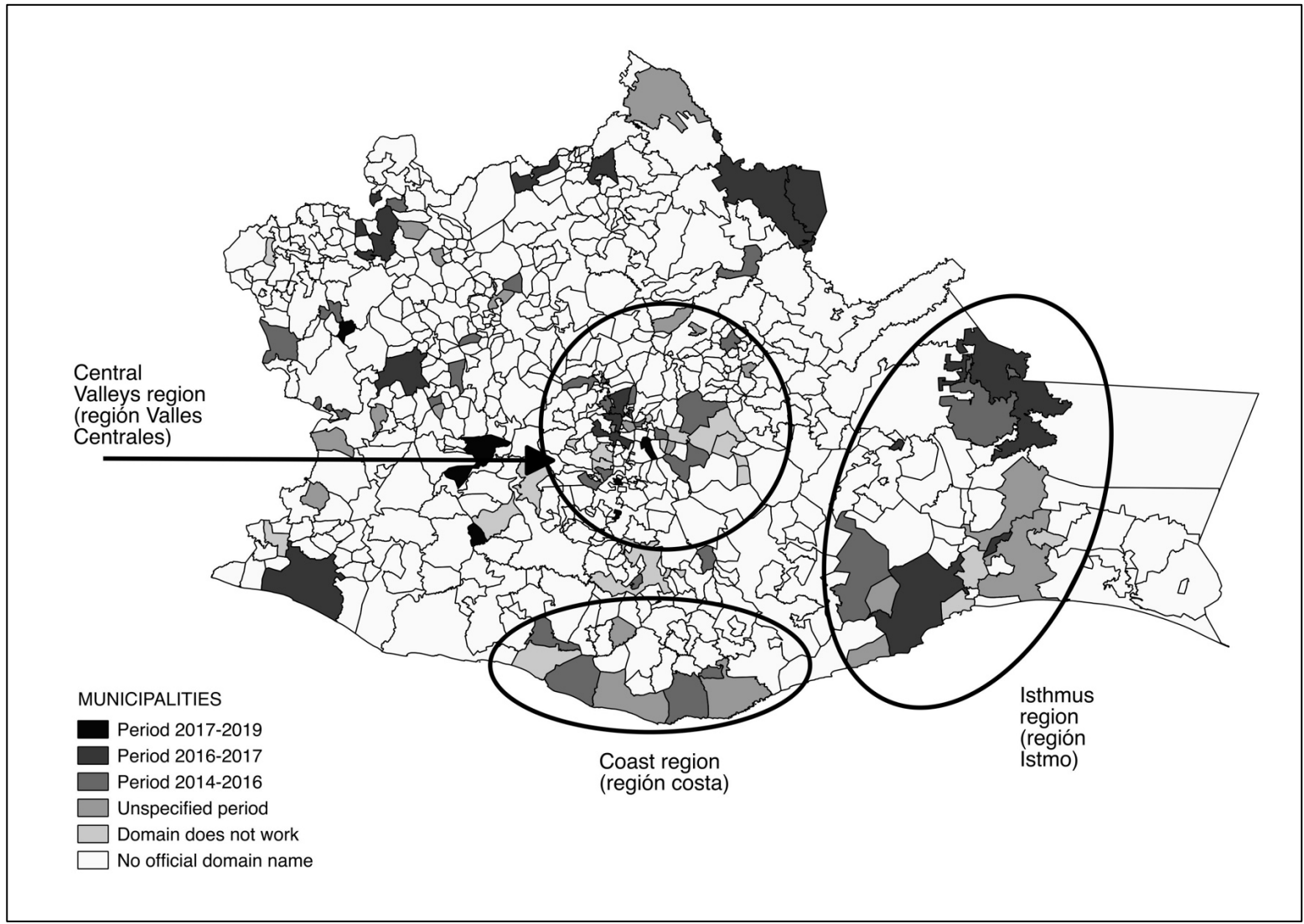

Source: own elaboration with information from the municipal websites (accessed: May, 24-28 and 31, 2017).

The websites heterogeneity can be seen in Table 5, which presents the titles (originally in Spanish) of the most frequent sections in the main menu at the website homepages: the most frequent titles are: *start* (53 cases, 13.6%), *transparency* (52, 13.3%), *contact* (30, 7.7%), *government* (19, 4.9%), *municipality* (17, 4.3%), and *news* (16, 4.1%). Other 35 diverse titles present frequencies between 2 and 16. Finally, there exist other 81 diverse titles with one case of each (this number is greater than 68 because more than one of these titles can occur in one single website).

**Table 5.** Titles (originally in Spanish) of the most frequent sections in the main menu of the municipal websites in the State of Oaxaca in late May, 2017.

| Section title | Number of websites | % | Section title | Number of websites | % |
|---|---|---|---|---|---|
| Start (*inicio*) | 53 | 13.6 % | Municipal family services (*DIF municipal*) | 3 | 0.8% |
| Transparency (*transparencia*) | 52 | 13.3 % | Directory (*directorio*) | 3 | 0.8% |
| Contact (*contacto*) | 30 | 7.7% | The municipality (*el municipio*) | 3 | 0.8% |
| Government (*gobierno*) | 19 | 4.9% | Councils (*regidurías*) | 3 | 0.8% |
| Municipality (*municipio*) | 17 | 4.3% | Services (*servicios*) | 3 | 0.8% |
| News (*noticias*) | 16 | 4.1% | Videos | 3 | 0.8% |
| Procedures and services (*trámites y servicios*) | 8 | 2.0% | Family activities (*actividades familiares*) | 2 | 0.5% |
| Press (*prensa*) | 7 | 1.8% | City council (*cabildo*) | 2 | 0.5% |
| City council (*Ayuntamiento*) | 6 | 1.5% | Mail (*correo*) | 2 | 0.5% |
| Works (*obras*) | 6 | 1.5% | City agencies (*dependencias*) | 2 | 0.5% |
| Transparency 2014 (*transparencia 2014*) | 6 | 1.5% | Citizen link (*enlace ciudadano*) | 2 | 0.5% |
| Transparency 2015 (*transparencia 2015*) | 6 | 1.5% | Events (*eventos*) | 2 | 0.5% |
| Transparency 2016 (*transparencia 2016*) | 6 | 1.5% | Homepage (*home*) | 2 | 0.5% |
| Galería (gallery) | 5 | 1.3% | Transparency reports (*informes de transparencia*) | 2 | 0.5% |
| City council (*H. Ayuntamiento*) | 5 | 1.3% | Public work (*obra pública*) | 2 | 0.5% |
| Public works (*obras públicas*) | 5 | 1.3% | Municipal development plan (*plan municipal de desarrollo*) | 2 | 0.5% |
| Your municipality (*tu municipio*) | 5 | 1.3% | Treasury (*tesorería*) | 2 | 0.5% |
| History (*history*) | 4 | 1.0% | Procedures (*trámites*) | 2 | 0.5% |
| Location (*ubicación*) | 4 | 1.0% | Tourism (*turismo*) | 2 | 0.5% |
| Activities (*actividades*) | 3 | 0.8% | Other titles (one case of each one) | 81* | 20.7% |
| Contracts (*contratos*) | 3 | 0.8% | **Average** | **9.5** | |
| | | | **Standard deviation** | **7.4** | |

Source: own elaboration using information from the municipal websites (accessed: May 24-28 and 31, 2017). *This number is greater than 68 because more than one of these titles can occur in one single website.

## 6.2. Technological feasibility for automating the classification of evolution development level

An exploration of the technological feasibility of our approach to automate the classification of the municipal websites is described below. Four websites are conventionally selected. There is one website belonging to each of the four levels. Their HTML source code is manually analyzed to search for cue words/phrases associated to each level. Figures 2 through 5 present the corresponding source code and a collection of cue words/phrases that are identified.

**Figure 2.** Example of a frequent cue word in the source code of a participation level website (municipality of Cajeme, Son.): *participativa* (participative).

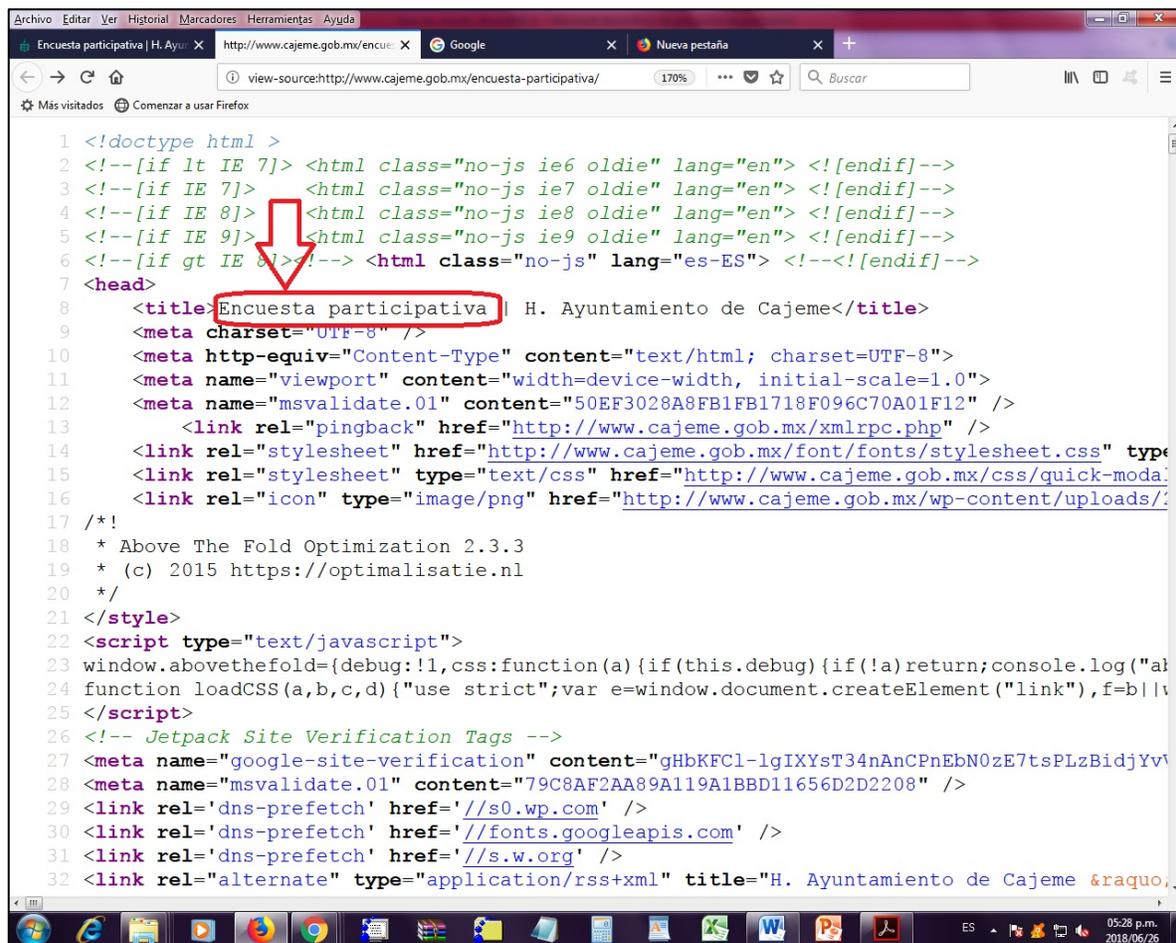

**Source:** http://www.cajeme.gob.mx/encuesta-participativa/ (accessed: June 26, 2018).

**Figure 3.** Example of a frequent cue phrase in the source code of a *transaction* level website (municipality of Merida, Yuc): *consulta y pago de predial* (consultation and payment of property tax).

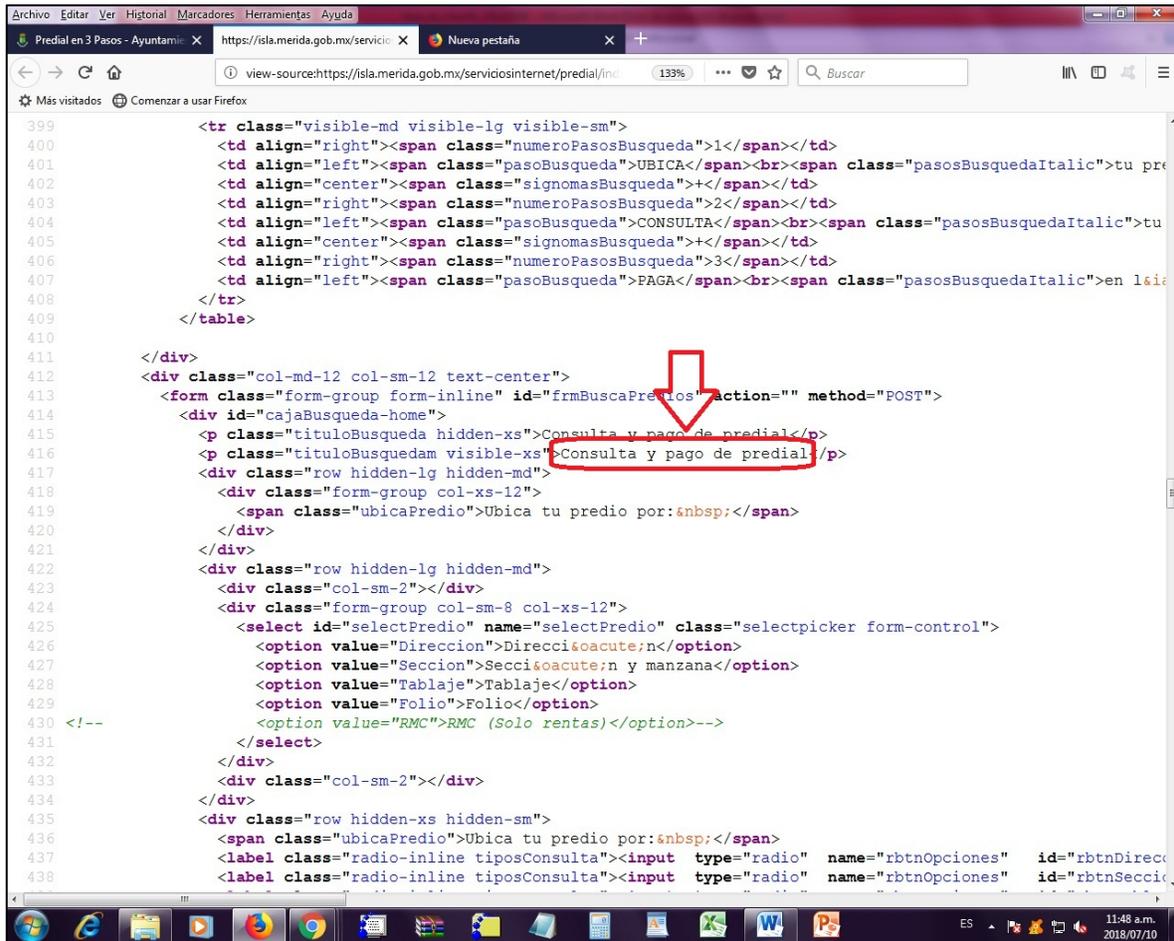

**Source:** https://isla.merida.gob.mx/serviciosinternet/predial/index.phpx (accessed: July 10, 2018).

**Figure 4.** Example of a frequent cue word in the source code of an *interaction* level website (municipality of Cadereyta, NL): *consulta tu predial* (consult your property tax).

**Source:** https://cadereyta.gob.mx/tramites-y-servicios/ (accessed: July 10, 2018).

**Figure 5.** Example of a frequent cue word in the source code of an information level website (municipality of Santa María Huatulco, Oax.): *transparencia* (transparency).

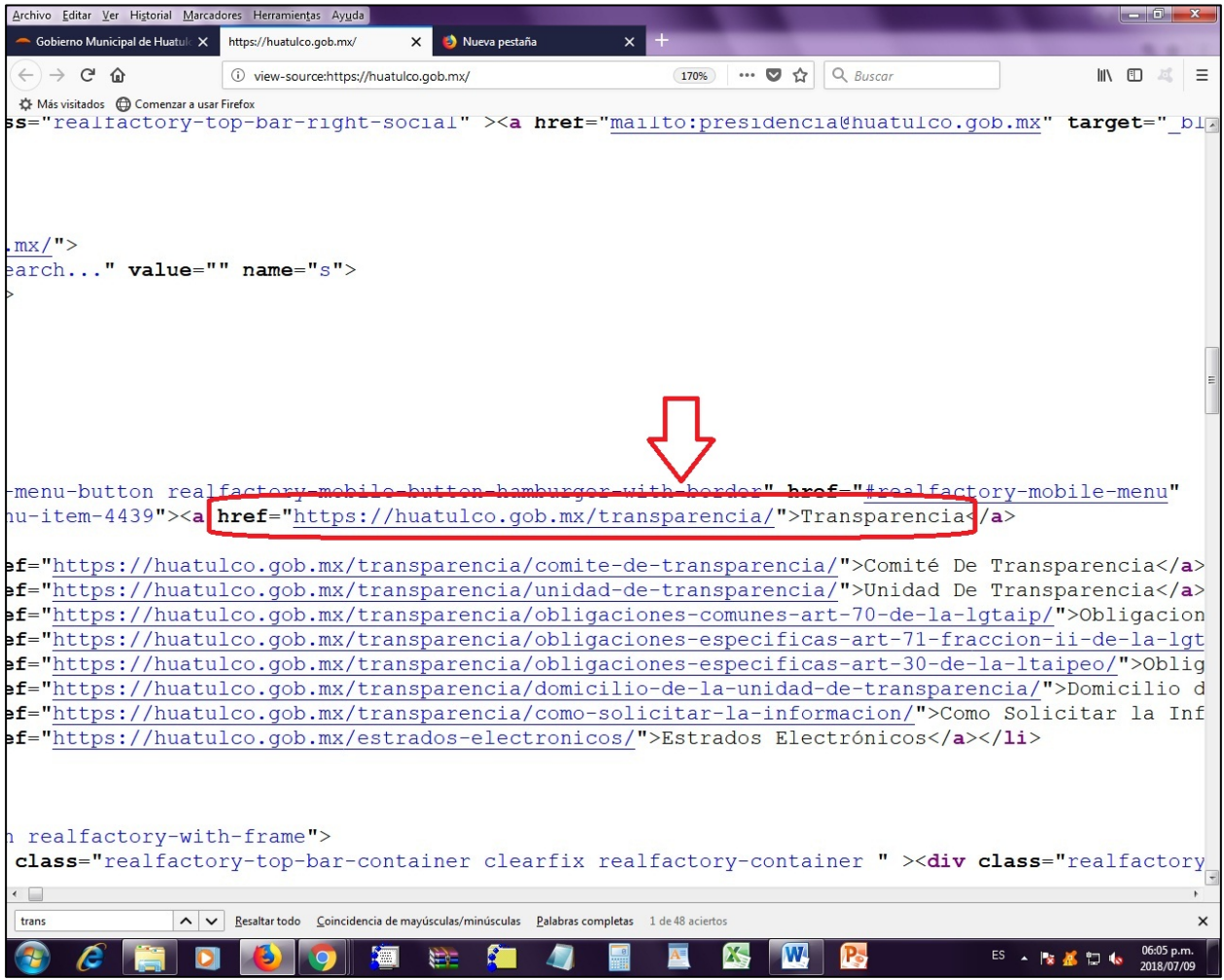

**Source:** https://huatulco.gob.mx/transparencia/ (accessed: July 09, 2018).

### 6.3. Discussion

The proposed method aims to be comprehensive and has the potential to yield a full list of existing websites because, although it is mainly based on automatic searching of the Mexican municipal websites, semi-automatic, and manual searching are also performed if the automatic searching is not successful. The only baseline for determining the completeness of this list is the list of all municipality names published by INEGI. The main source of error is the fact that the number of municipal governments that do not have an official website at a given time is unknown. Thus, a major problem in the process is deciding if a municipal website should be assumed as non-existing if it cannot be found after automatic, semi-automatic and manual searching. In turn, a potential solution to this problem is defining a conventional *stop condition* when this occurs; for instance, assuming the website as non-existing if a manual searching could not find it.

## 7. Software requirements specifications to automate the creation, updating and analysis of an online repository of municipal website replicas

Researchers, practitioners and users of e-government in Mexico need an information technology solution that allows them to know which municipalities have an operating and valid e-government website at a given time. Also, specialized users can benefit from automatic features to identify the evolution development level of these websites. One of the potential solutions is to design and implement an online system that automatically produces and updates a directory of municipal websites and a digital repository of replicas of them. On the basis of the manual search and statistical analyses of the municipal websites that are described above, a set of functional requirements can be specified for the subsequent development of a system to produce, preserve, update, and analyze a digital repository of replicas of these websites. The repository should be publicly available at the WWW and preserved indefinitely. Its purpose is to provide users with information resources to perform diverse tasks on collection, organization, information extraction, statistical analyses, and characterization with both cross-sectional and longitudinal approaches. A highly valuable guide to specify the software requirements is (IEEE, 1998).

Our review on scientific literature (see *Section 4*) suggests potential analyses and characterizations that can be performed on the Mexican websites as well as techniques and software tools for web scraping and parsing, and automatic classification of websites. On this basis, the directory and repository system should:

i. Store a replica of the official list from INAFED containing the official domain names of the municipal websites. Previously, the list should be manually downloaded from the INAFED website and properly formatted to be fed into the system.

ii. Automatically verify that every domain name in the list corresponds to a working website and produce a verified list (i.e. a directory) as an output file, including the operating status as one of its data items. The directory should be implemented as a database (either relational, CSV or other type). The date of access to every website should be added in the directory.

iii. Using the directory as an input, associate every domain name to its corresponding municipality ID defined by INEGI and add this data to the directory. A table with the municipalities names and their INEGI ID should be previously obtained using manual processes.

iv. Using the directory as an input, identify the period of the municipal government that produced the contents of every website. Web scraping tools can be used for this purpose. Add the government period to the directory.

v. Using the directory as an input, automatically identify the name and location country of the web hosting provider for every website and add this information to the directory.

vi. Allow users to consult the directory and export any data from it as a CSV file.

vii. Allow users to specify a maximum number of depth levels in the websites, a maximum number of files, and a maximum file size, for downloading processes, so that the system does not download too many files (or too large files). At least, the main homepage (usually entitled as *Start*) of every website should be downloaded.

viii. Allow users to specify the types of files to be downloaded (e.g. html, php, jsp, css, jpg, png, pdf, etc.) into the website replicas repository, including the option to download all type files.

ix. Using the directory as an input, automatically download a copy of the web page files from every municipal website considering the maximum number of depth levels in the websites or the maximum number of files to be downloaded that are specified by user. The downloaded files should be organized so that they facilitate information extraction and statistical analyses.

x. Using the directory and the websites repository as inputs, automatically produce a list of the major sections in every website by consulting the section titles in the main homepages. This list should be implemented as a database (either relational, CSV or other type).

xi. Allow users to consult the section titles and to export any data from them as CSV files.

xii. Automatically produce Pareto analyses for: operating status of the websites, government periods, hosting providers and their location countries, frequency of section titles, number of section titles in the websites, and number of websites per number of sections.

xiii. Allow users to consult the Pareto analyses as tables and statistical charts.

xiv. Automatically produce choropleth maps of the States showing the status of their municipal websites and the corresponding government periods.

xv. Automatically determine the evolution development level of the websites by using known techniques and tools for web scraping and parsing, and for web page classification (*e.g.* the presented in the literature review in Table 2).

## 8. Conclusion

This article has addressed a problem in the electronic government and computer science disciplines: the identification, quantitative analysis, and characterization of the e-government websites of the Mexican municipalities. The article has explored the need and the technological feasibility for a digital repository of replicas of these websites as a means to facilitate their consultation, analysis, and characterization. Our approach has been quantitative, descriptive and exploratory.

Research on websites of the Mexican municipalities is highly scarce. Due to the relatively large number of municipalities (more then 2,450) and the scarcity of official sources of information on these websites (such as the Mexican Institute for the Federalism and Municipal Development), the searching and analysis are highly time-consuming. In addition, automating these tasks is an interesting challenge from the computer science and engineering perspectives. Therefore, there is a need for methodologies and automatic tools for these purposes. This need is justified due to the potential social impacts of e-government to promote transparency, accountability, and civic engagement in municipalities.

Our review on scientific literature shows that municipal websites in other countries are a highly frequent subject matter in the e-government discipline and that the most frequently studied aspects are: evolution development level, usability, compliance with legal specifications, among others. In turn, the literature review on techniques and tools for web scraping shows that there exist a number of recent free software tools, for instance in *R* and *Python* languages, that can be easily harnessed for automate the classification of the evolution development level of the websites on the basis of cue words and phrases.

With this in mind, a methodology for collection and analysis of these websites as well as for classifying their evolution development level has been proposed. This is organized into the steps enumerated below: obtain a preliminary list of municipal websites that is available at INAFED; using this list as an input, verify the validity of the websites by a semi-automatic process applying a web crawling free software tool (for instance, J-Spider, or a convenient package in *R* or *Python* languages); identify the web hosting provider of every website (e.g. by using the *whoishostingthis.com* service or other similar); identify the major section titles in every website by analyzing the main menu in its homepage (e.g. by using *Google WebScraper*, *www.octoparse.com*, or *R* or *Python* packages); produce Pareto analyses of each of the enumerated characteristics, and produce a choropleth map of the municipalities showing the status of their websites. In order to automate the classification of evolution development level of the websites, our proposal involves identifying cue words or cue phrases that can be parsed in the websites HTML source code by using web scraping, and parsing tools already implemented and freely available as *R* and *Python* packages. Therefore, a key requirement to automate the classification is producing a comprehensive enough list of cue words/phrases that are most frequently associated to each evolution development level.

Statistical evidence supporting our proposal is provided, containing preliminary results about the existing e-government websites of the 570 municipalities in the State of Oaxaca, Mexico, in late May, 2017. The results in this research consisted in a series of statistical tables and a map presenting patterns of the websites of those municipalities. The most remarkable patterns discovered involve: municipalities that have or not an official (*.gob.mx*) web domain name, and from these, those that are working; validity of the websites, i.e. how many correspond to current (in 2017) municipal governments, how many correspond to finished governments (period 2014-2016), and how many do not specify their government period. The map shows that most of the municipalities in the State do not have an official web domain name, and the municipalities that do have it are mostly located in three geographic regions of the State: Central Valleys (*Valles Centrales*), Coast (*Costa*), and Isthmus (*Istmo*). Regarding the websites contents, most of them contain between 4 and 7 sections, so these seem to be the numbers of sections preferred by municipal authorities. There are extreme cases: one website with 14 sections, and one with 2. The 5 most frequent section titles are: *Start* (13.6%), *Transparency* (13.3%), *Contact* (7.7%), *Government* (4.9%), and *Municipality* (4.3%). There are other 35 section titles with frequencies between 2 and 16. Finally, there are other 81 section titles with frequencies equal to 1, involving that more than one of these titles can exist in one single website.

As evidences on the technological feasibility of automating the classification of evolution development level of websites, we have offered our review of scientific literature on web scraping techniques and tools, and our empirical observation of selected actual municipal websites. The literature review shows that the automatic classification of websites by using parsing with cue words or cue phrases is a process implemented and used on diverse knowledge fields since a number of years ago. In turn, our empirical observation shows that particular sets of cue words and phrases corresponding to each evolution development level can be identified in actual municipal websites.

By contrasting this research to previous works in the field, our most relevant contributions are: first, the identification of the need for producing and keeping updated a list (directory) of valid official Mexican municipal websites, and a digital repository of website replicas. Both the directory and the repository should be publicly available on the Web and will offer the possibility to analyze, mainly: 1) which and how many municipalities have a valid and working e-government website at a given time, 2) what are the section titles their contents are organized into, 3) what is the evolution development level of each website, and 4) how the websites change along time. A second contribution is a methodology to perform searching and statistical analyses of the websites. The web data mining discipline can offer methods and software tools to automate or semi-automate these tasks. The third contribution is a preliminary simple technique to automate the classification of evolution development level of the websites. Finally, a fourth contribution is a set of basic software requirements specifications to implement the repository. Our claims are that a digital repository of replicas of Mexican municipal websites is a necessary and useful resource for theoretical and practical purposes, and that it is technologically feasible on the basis of available techniques and free software tools for statistics and web scraping. Cue words and phrases in the websites textual contents can be used to automate the classification process. The overall significance of the findings consists in that these are useful for creating a highly valuable information resource (i.e. the repository) for studying and promoting the e-government in Mexican municipalities. Also, these findings can be harnessed to implement similar repositories in other countries for academic, government or business purposes.

Future research work on this topic should complement, deepen and refine the informational and functional requirements enumerated above to design and fully implement the directory and the repository. These two information resources will be highly useful for research purposes in the public administration, e-government, software engineering, and computer science fields. Suitable updating frequencies for the directory and the repository should be defined and this information should be introduced into the system as settings by an administrator user. By this way, the system can automatically perform the updating of the directory and the insertion of new replicas in the repository corresponding to recent versions of the website pages. Besides, machine learning techniques can be harnessed to discover diverse interesting and relevant patterns; for instance, classification tree algorithms can be used to discover the patterns implicit in tags and text contents in the web pages.

Practical applications of the preliminary statistical outcomes can be, for instance: promoting and facilitating that all Mexican municipalities to obtain official domain names (*.gob.mx*) to create their websites and that these can use the secured *http* protocol (*https*); offering information technology consulting to municipal governments so that they can create and administrate their websites; creating technical standards or guidelines for design of municipal websites; promoting that municipal governments with experience in creation and administration of their websites share experiences with their neighbor municipalities.

At a later stage, this system can also allow the extraction of useful information from the websites to produce specialized databases or directories, to verify the compliance with legal specifications, etc. Finally, this analysis methodology and the software requirements specifications can be adapted to be applied on municipal websites in other countries, so this research could be the initial stage of an effort to produce long term longitudinal studies worldwide.